\documentstyle[prc,preprint,tighten,aps]{revtex}
\begin{document}
\draft
\title{Off-shell effects in the energy dependence of the 
$\bbox{^7}$Be$(\bbox{p,\gamma})$$\bbox{^8}$B astrophysical 
$\bbox{S}$ factor}  
\author{Attila Cs\'ot\'o \thanks{csoto@qmc.lanl.gov}} 
\address{Theoretical Division, Los Alamos National Laboratory, 
Los Alamos,  NM 87545, USA}
\date{28 October, 1996}

\maketitle

\begin{abstract}
\noindent
I show that off-shell effects, like antisymmetrization and
$^7$Be distortions, can significantly influence the energy
dependence of the nonresonant $^7$Be($p,\gamma$)$^8$B
astrophysical $S$ factor at higher energies. The proper
treatment of these effects results in a vitrually flat $E1$
component of the $S$ factor at $E_{cm}=0.3-1.5$ MeV energies  
in the present eight-body model. The energy dependence of the 
nonresonant $S$ factor, predicted by the present model, is 
in agreement with the low-energy direct capture data and the 
existing high-energy Coulomb dissociation data. Irrespective 
of whether or not the present energy dependence is correct, 
off-shell effects can cause 15--20\% changes in the value 
of $S(0)$ extrapolated from high-energy ($E_{cm}>0.7$ MeV) 
data.
\end{abstract}

\narrowtext

The most uncertain nuclear input parameter in Standard Solar
Models, which are used to calculate the solar neutrino fluxes
on Earth, is the low-energy $^7$Be($p,\gamma$)$^8$B radiative
capture cross section. This reaction produces $^8$B in the
Sun, whose $\beta^+$ decay is the main source of the 
high-energy solar neutrinos. Many present (Homestake
\cite{Homestake}, Kamiokande \cite{Kamiokande}, Superkamiokande
\cite{super}) and future (SNO \cite{SNO}) solar neutrino
detectors are sensitive mainly or exclusively to the $^8$B
neutrinos. The theoretically predicted $^8$B neutrino flux is
proportional to the very low-energy ($E_{cm}=20$ keV; $E_{cm}$
is the $^7$Be--$p$ scattering energy in the CM frame) 
$^7$Be($p,\gamma$)$^8$B strophysical $S$ factor, $S_{17}$.
Thus, precise knowledge of $S_{17}$(20 keV) is crucial to
understand the solar neutrino problem \cite{Bahcall}.

Currently there is a considerable confusion concerning the
value of $S_{17}(0)$. The six direct capture measurements
performed to date give $S_{17}(0)$ between 15 eVb and 40 eVb,
with a weighted average of $22.2\pm 2.3$ eVb
\cite{Johnson}. All these $S_{17}(0)$ values were determined
by extrapolating from higher energies ($E>100$ keV) where the
experiments are feasible. The two lowest energy measurements
by Kavanagh {\it et al.} \cite{Kavanagh} and by Filippone {\it
et al.} \cite{Filippone} disagree by 25\%.

Recently the Coulomb dissociation process was suggested as a
promising new method to measure the cross sections of
astrophysical processes \cite{Baur}. Based on this technique,
the $^7$Be($p,\gamma$)$^8$B cross section was studied recently
by Motobayashi {\it et al.} using a radioactive $^8$B beam at
RIKEN \cite{Motobayashi}. The extrapolated zero energy $S$
factor is $S_{17}(0)=16.7\pm 3.2$ eVb \cite{Motobayashi}, or 
$S_{17}(0)=15.5\pm 2.8$ eVb \cite{Shyam} using $s$-wave
\cite{Tombrello}, or $s+d$-wave \cite{Typel} extrapolations,
respectively. The theoretical implications of this experiment
were studied in several papers, see e.g.\ 
\cite{Shyam,Typel,LS,Gai,LSrep,Bertulani,Esbensen,Esbensen1,Typel2}. 
Although the role of the $E2$ transitions 
in this process is heavily debated 
\cite{LS,Gai,LSrep,Bertulani}, it appears that, although 
there is substantial $E2$ strength present, the Coulomb 
dissociation measurement of \cite{Motobayashi} gives, in a 
good approximation, the $E1$ component of the cross section 
at the whole measured energy range \cite{Esbensen1,Iwasa}. 
The $M1$ transition is suppressed in \cite{Motobayashi} 
because of the low virtual $M1$ photon flux. 

New $^8$B Coulomb dissociation experiments to study the
$^7$Be($p,\gamma$)$^8$B cross section have been 
performed at RIKEN \cite{Gaip} and at MSU NSCL \cite{Austin}, 
and are being planned at GSI \cite{Gaip}. 

The theoretical predictions for $S_{17}(0)$ also have a huge
uncertainty, as the various models give values between 16 eVb
and 30 eVb \cite{Langanke}. The energy dependence of the $S$
factor was first studied by Christy and Duck \cite{Christy} 
and by Tombrello \cite{Tombrello} using only $s$-waves for the 
scattering states. Later the importance of $d$-waves was 
emphasized by Robertson \cite{Robertson}, and especially by 
Barker \cite{Barker,Barker_d}. Despite the large differences
among the various theoretical predictions, it is now a common
belief that the energy dependence of the nonresonant $S$ 
factor is known, and the differences come only from the
absolute normalization. I point out in this paper that
unfortunately this claim is not substantiated. Off-shell
effects can significantly influence the energy dependence of
the $S$ factor and thus also the extrapolation procedure, and
the extracted value of $S_{17}(0)$.

The importance of off-shell effects was recently emphasized in
the mirror reaction, $^7$Li($n,\gamma$)$^8$Li \cite{Brown}.
The thermal cross section of this reaction was used by Barker
to argue that potential models tend to overestimate
$S_{17}(0)$ \cite{Barker}. In Ref.\ \cite{Brown} it was
pointed out that although the on-shell properties of the wave
functions are well determined (e.g.\ $^7$Li+$p$ scattering
lengths), a change in the unknown off-shell (internal) 
structures of the $^7$Li+$p$ scattering wave functions can
drastically influence the thermal neutron capture cross
section, while the same change leaves $S_{17}(0)$ virtually
unaffected.

In order to reliably extrapolate the direct capture $S_{17}$ 
factors to zero energy, the energy dependence of the
nonresonant $E1$ term should be known. The $E2$ cross section
is not expected to play a role in the $E_{cm}=0-3$ MeV 
energy range \cite{Kim}. The low-energy ($E_{cm}<300-400$ keV) 
direct capture measurements record probably pure $E1$ cross
section, while at and above the $1^+$ resonance at
$E_{cm}=632$ keV they see a mixture of $E1$ and $M1$
transitions. As I mentioned above, the $^8$B Coulomb
dissociation measurement \cite{Motobayashi}, in good 
approximation, gives the $E1$ component of the cross section 
at 0--2 MeV energies. 

In the present work I concentrate on the $E1$ transition, 
which connects the $J^\pi$=$1^-$, $2^-$, and $3^-$ $^7$Be+$p$ 
scattering states with the $2^+$ ground state of $^8$B. 
Using the eight-body model of Ref.\ \cite{Be7p} I show that, 
although not as much as in the case of
$^7$Li($p,\gamma$)$^8$Li, the internal structure of the wave
functions can significantly influence the energy dependence 
of the $E1$ component of the $S$ factor.

A $^4$He+$^3$He+$p$ three-cluster eight-body RGM approach 
is used for both the $^8$B ground state and the $^7$Be+$p$ 
scattering states. The eight-body wave functions have the form
\begin{equation}
\Psi=\sum_{I_7,I,l_2}\sum_{i=1}^{N_7}
{\cal A}\Bigg \{\bigg [ \Big [\Phi^p_s\Phi^{^7{\rm
Be},i}_{I_7} \Big ]_I
\chi_{l_2}^i(\bbox{\rho }_2) \bigg ]_{JM} \Bigg \},
\label{wf}
\end{equation}
where ${\cal A}$ is the intercluster antisymmetrizer,
$\bbox{\rho}_2$ and $l_2$ is the relative coordinate and
relative angular momentum between $^7$Be and $p$,
respectively, $s$ and $I_7$ is the spin of the proton and
$^7$Be, respectively, $I$ is the channel spin, and [...]
denotes angular momentum coupling. While $\Phi^p$ is a proton
spin-isospin eigenstate, the antisymmetrized ground state 
($i=1$) and continuum excited distorted states ($i>1$) of 
$^7$Be are represented by the wave functions
\begin{equation}
\Phi^{^7{\rm Be},i}_{I_7}=\sum_{j=1}^{N_7}c_{ij}\sum_{l_1} 
{\cal A}\left \{\left [ \left [\Phi^\alpha\Phi^h \right
]_{\frac{1}{2}}\Gamma_{l_1}^j(\bbox{\rho }_1) \right ]_{I_7M_7} 
\right \}.
\end{equation}
Here ${\cal A}$ is the intercluster antisymmetrizer between
$\alpha$ and $h$, $\Phi^\alpha$ and $\Phi^h$ are
translationally invariant harmonic oscillator shell model
states ($\alpha$=$^4$He, $h$=$^3$He), $\bbox{\rho}_1$ is
the relative coordinate between $\alpha$ and $h$, $l_1$ is the
$\alpha -h$ relative angular momentum, and
$\Gamma_{l_1}^j(\bbox{\rho_1})$ is a Gaussian function with a 
width of $\gamma_j$. The $c_{ij}$ parameters are determined
from a variational principle for the $^7$Be energy. In the
$^8$B ground state and $^7$Be+$p$ scattering states $N_7$=6,
$l_1$=1, $I_7$=3/2,1/2, $I$=1,2, and $l_2$=0,1,2 are used. 

Putting (\ref{wf}) into the eight-body Schr\"odinger equation,
which contains the Minnesota nucleon-nucleon interaction
\cite{Be7p}, we arrive at an equation for the unknown relative
motion functions $\chi_{l_2}^i(\bbox{\rho }_2)$. These
relative motion functions are determined from a variational
Siegert method \cite{Fiebig} for the $2^+$ bound groundstate
of $^8$B, and from the Kohn-Hulth\'en variational method
\cite{Kamimura} for the $1^-$, $2^-$, and $3^-$ $^7$Be+$p$
scattering states \cite{Be7p}. 

In (\ref{wf}) we neglect the $h(\alpha p)$ and $\alpha (hp)$
type three-body rearrangement channels. In \cite{Be7p} it was
shown that the present model (called $^7$Be+$p$ type model
space in \cite{Be7p}) and a model which contains these
rearrangement channels (called full model space in
\cite{Be7p}) lead to similar results, provided the subsystem
properties (e.g., the $^7$Be quadrupole moment) are reproduced
equally well. The presence of these rearrangement channels
would not change the conclusions of the present work. Further
details of the model, parameters, $N-N$ interaction, etc.\ can
be found in \cite{Be7p} and \cite{B8}.

The $E1$ $^7$Be$(p,\gamma$)$^8$B radiative capture
cross section is given by \cite{pert}
\begin{eqnarray}
\sigma (E_{cm}) &=& \sum_{J_i} {{1}\over{(2I_7+1)(2s+1)}}
{{16\pi}\over{3\hbar}} 
\left ({{E_\gamma}\over{\hbar c}}\right )^{3} \cr
&\times& \sum_{l_\omega,I_\omega}(2l_\omega+1)^{-1} 
\vert \langle \Psi^{J_f} \vert \vert {\cal
M}_1^E \vert \vert
\Psi^{J_i}_{l_\omega,I_\omega} \rangle \vert ^2,
\label{sigma}
\end{eqnarray}
where $I_7$ and $s$ are the spins of the colliding 
clusters, ${\cal M}_1^E$ is the electric dipole ($E1$)
transition operator, $\omega$ represents the entrance 
channel, $E_\gamma=E_{cm}+0.137\;$MeV is the photon energy, 
and $J_f$ and $J_i$ is the total spin of the final and initial 
state, respectively. The initial wave function 
$\Psi^{J_i}_{l_\omega,I_\omega}$ is a partial wave of a 
unit-flux scattering wave function. 

As far as off-shell properties are concerned, the main
differences between the present model and the potential
models, e.g.\ \cite{Xu}, are the following: (i) the presence
of the $^7$Be--$p$ antisymmetrization in (\ref{wf}); (ii) the
presence of the continuum excited $^7$Be distortion channels
in (\ref{wf}); (iii) the node positions in the scattering
states can be different. The node positions are determined by
nucleon exchange and antisymmetrization effects, so only the
microscopic models are in firm ground in this respect. All the
above differences have short range effects.

In Fig.\ 1 I show the $^7$Be($p,\gamma$)$^8$B astrophysical
$S$ factor coming from the present eight-body model (solid
line) together with the results of the direct capture
\cite{Kavanagh,Filippone,Vaughn,Parker} and Coulomb 
dissociation \cite{Motobayashi} experiments, and a typical 
$s+d$-wave potential model $S$ factor (dotted line) from 
\cite{Xu}. The $S$ factor, as a function of the $^7$Be--$p$
scattering energy, is defined as 
\begin{equation}
S(E_{cm})=\sigma (E_{cm})E_{cm}\exp \Big [2\pi\eta (E_{cm}) 
\Big ],
\end{equation}
where $\eta$ is the Sommerfeld parameter. 

One can see that the energy dependence of the $S$ factor in
the present model is rather different from that in a potential
model. The latter is often referred in the literature as the
``known energy dependence of the $S$ factor''. Note that the
present model predicts a virtually flat $S$ factor at
$E_{cm}=0.3-1.5$ MeV, despite the presence of $d$-waves in the
scattering wave functions. The two other nucleon-nucleon
forces used in Ref.\ \cite{Be7p} give very similar results,
except that the $S$ factor coming from the $V2$ force starts
to rise at around 1.0 MeV.

To show that the difference between the present model and the
potential model comes from the above-mentioned off-shell
effects, the antisymmetrization, which has the biggest effect,
is switched off in (\ref{sigma}). In other words we take the
relative motion wave functions $\chi$ {\it behind the
antisymmetrizer} in (\ref{wf}) and use them as if they were
coming from a potential model. Thus (\ref{sigma}) reduces to
one-dimensional integrals. The resulting $S$ factor is
shown by the dashed line in Fig.\ 1. Its energy dependence is
much closer to the energy dependence coming from the potential
model, although its slope is smaller. This is not surprising as
there still are off-shell differences between the two models, 
namely the $^7$Be distortion states in (\ref{wf}) and the 
different node positions.

Irrespective of the fact whether or not the solid line in
Fig.\ 1 represents the ``true'' energy dependence of the $S$
factor, differences in model predictions for the energy
dependence, at least as big as the difference between the
solid and dashed lines, can be expected from off-shell
effects. As these effects are completely missing in potential
models, their energy dependence cannot be accepted as ``the
known energy dependence of the $S$ factor''. The microscopic
models are also not free from defects, unfortunately. In some
cases, for instance, the description of $^7$Be in the
asymptotic region is unsatisfactory, in others the $^7$Li+$n$
scattering lengths (and supposedly also the $^7$Be+$p$ ones)
are wrong. The present model is also not perfect because,
e.g., as in \cite{Be7p} the scattering wave functions are
calculated from uncoupled-channel models. My test calculations
show that this has little effect on the results, nevertheless
an improvement would be desirable. What the present model does
not contain, but what can prove to be important, are the
dynamical degrees of freedom beyond the three-cluster model.
Hopefully fully dynamical ab initio \cite{Joe} eight-body 
calculations for this reaction will be feasible in the future.

I would like to emphasize again, that whether or not the
present model will prove to be a good description of the
$^7$Be($p,\gamma$)$^8$B reaction, Fig.\ 1 shows the amount of
change in the $S$ factor energy dependence one can expect
from off-shell effects.

Finally, I would like to point out that the energy dependence
predicted by the present model (solid line in Fig.\ 1) is in
agreement with the low-energy direct capture data, and the
high-energy Coulomb dissociation measurements. I note, that
in order to reproduce the {\em absolute normalization} of the 
data, the different experiments would require different 
renormalizations of the theoretical curve. There is no physical 
reason for such renormalizations in the present model; the 
zero energy $S$ factor is $S_{17}(0)=25$ eVb. The energy 
dependence of the solid line is in disagreement with the 
direct capture results above 0.7 MeV, if it is forced to 
fit the low-energy data points. In some direct
capture experiments, e.g.\ in Ref.\ \cite{Filippone}, the 
parameters of the 632 keV resonance were determined by
fitting the data with a Breit-Wigner form (with energy 
dependent width) after a nonresonant part was removed using 
the energy dependence of a potential model, e.g.\ 
\cite{Tombrello}. Assuming a different nonresonant $E1$ 
energy dependence, the resonance parameters could slightly 
be different, modifying the experimental $M1$ contribution 
at 0.7--1.0 MeV. I also note, that \cite{Kim} and 
\cite{Baye} find a $1^+$ state in $^8$B at around 1.3--1.4 
MeV, and there is a $3^+$ state at 2.2 MeV 
\cite{Ajzenberg}, which can cause $M1$ bumps at these 
energies (see e.g.\ \cite{Kim}). The present model also 
predicts a second low-lying $1^+$ state in $^8$B above the 
632 keV resonance. Thus, as the direct capture results 
come from an $E1+M1$ mixture at higher energies, the 
disentanglement of these components may prove to be 
difficult.

In summary, using an eight-body model of the
$^7$Be($p,\gamma$)$^8$B reaction, I have shown that off-shell
effects can considerably change the energy dependence of the
of the astrophysical $S_{17}$ factor. The model predicts an
$S$ factor which is virtually flat in the 0.3--1.5 MeV
energy range, in strong contrast to the ``standard'' potential
model energy dependence. I have demonstrated that using the
$^7$Be--$p$ relative motion wave functions, which are behind
the antisymmetrizer in the microscopic model, as if they came
from a potential model, the energy dependence of the $S$
factor becomes similar to the ``standard'' shape. Thus, the
present unusual energy dependence of the $S$ factor is caused
mainly by the antisymmetrization, the biggest off-shell effect.
I have argued that because the direct capture experiments
contain a mixture of $E1$ and $M1$ transitions for 
$E_{cm}>0.5$ MeV, the knowledge of the true energy dependence 
of the $E1$ $S$ factor is crucial to the extrapolation of 
the high-energy ($E_{cm}>0.7$ MeV) data to zero energy.

The energy dependence predicted by the present microscopic
model is in agreement with the low-energy direct capture data
and the high-energy Coulomb dissociation results. However, I
must point out that this agreement might only be fortuitous, 
and the present energy dependence of the $S$ factor might 
prove to be incorrect, using a larger eight-body model space. 
Nevertheless, off-shell effects, at least as big as in 
Fig.\ 1, should be expected in the $^7$Be($p,\gamma$)$^8$B 
reaction. This can change the zero energy $S_{17}$, 
extrapolated from high-energy data, by as much as 15--20\%.

Note: In a very recent preprint \cite{Nunes} it was pointed
out that in the potential model the energy dependence of the
$E1$ $S$ factor is strongly influenced by short range effects,
e.g., by $^7$Be core deformations.

\mbox{\ }

This work was performed under the auspices of the US 
Department of Energy. I would like to thank Sam Austin for
useful discussions and Moshe Gai for sending me the tabulated
values of the experimental $S$ factors (compiled by Calvin
Johnson). Early stages of this work were done at Caltech W.\
K.\ Kellogg Radiation Laboratory and at the Michigan State
University National Superconducting Cyclotron Laboratory,
supported by NSF Grants.\ PHY94-15574, PHY94-12818, 
PHY92-53505, and PHY94-03666. Support from OTKA Grant F019701
is also acknowledged.

\mediumtext
\begin{figure}
\caption{Astrophysical $S$ factor for the
$^7$Be($p,\gamma$)$^8$B reaction. The experimental data are
from the direct capture measurements
\protect\cite{Kavanagh,Filippone,Vaughn,Parker} and from the 
Coulomb dissociation of $^8$B \protect\cite{Motobayashi}. 
The solid and dashed lines are the $E1$ components of the 
$S$ factors in the present eight-body model with and without 
antisymmetrization in the electromagnetic transition matrix 
element, respectively. The dotted line is the result of the 
potential model \protect\cite{Xu}.}
\label{fig1}
\end{figure}

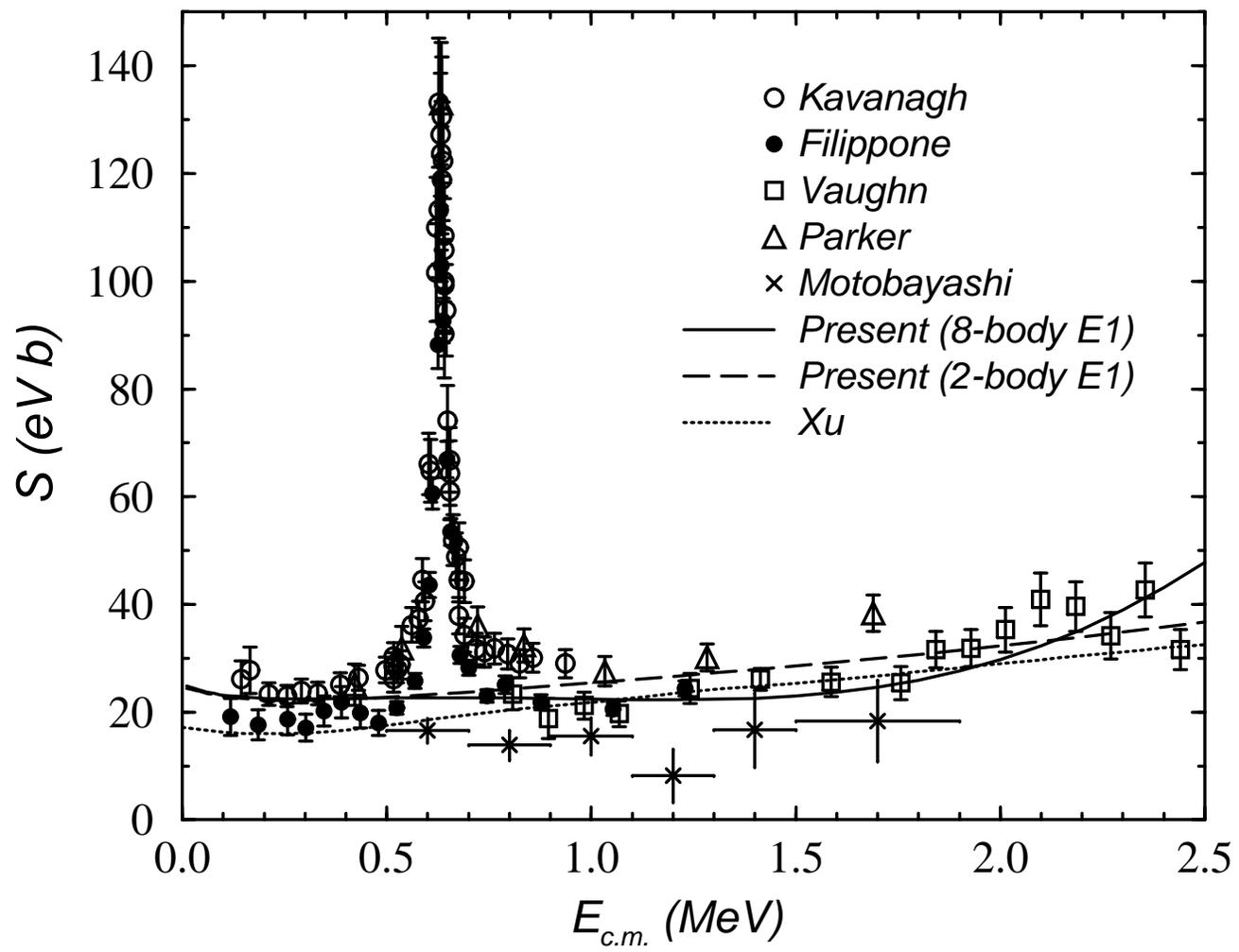
\end{document}